\documentclass{PoS}
\usepackage{amsmath}
\usepackage{xcolor}
\definecolor{orange}{rgb}{1,0.5,0}

\title{Monte Carlo simulation of $\phi^4_2$ and $O(N)\phi^4_3$ theories}

\ShortTitle{MC simulation of $O(N)\phi^4$ theory}

\author{\speaker{Barbara De Palma} $^a$ and Marco Guagnelli $^b$\\
        \llap{$^a$} Universit\`a degli studi di Pavia and INFN\\
        \llap{$^b$} INFN \\
        E-mail: \email{barbara.depalma01@universitadipavia.it}, \email{marco.guagnelli@pv.infn.it}}

\abstract{We report lattice simulations of $\phi^4_2$ and $O(N)\,\phi^4$ models, performed by
means of a Monte Carlo method based on the all-order strong coupling expansion (worm
algorithm). The investigation of the non-perturbative features of the $\phi^4$
continuum limit in two dimensions lead us to the result $g/\mu^2 = 11.15 \pm
0.06_{stat} \pm 0.03_{syst}$ for the critical coupling. Furthermore we present preliminary results for the three-dimensional $O(2)\phi^4\,$ model using the worm algorithm with the extention to $O(N)\phi^4\,$ in $D$ dimensions.}

\FullConference{34th annual International Symposium on Lattice Field Theory\\
		24-30 July 2016\\
		University of Southampton, UK}

\begin{document}

\section{Introduction}
The $\phi^4$ theory plays a pedagogical role in the study of quantum field theories. Despite the simplicity of the model there is still interest in the investigation of its main features by means of numerical and theoretical tools.
As often happens, significant results in the theoretical study goes hand by hand with the efficiency of algorithms and  computational power. 
From the point of view of simulations on the lattice it is fundamental to obtain precise results in the continuum limit, i.e. when the lattice spacing $a$ goes to zero and the lattice is removed. In this limit the correlation length $\xi$ diverges and the problem of \emph{critical slowing down}, in which the autocorrelation time $\tau$ grows like a power of $\xi$, occurs. For real simulations on a lattice of linear size $L$, $\xi\sim L$ and $\tau\sim L^z$, where $z$ is the dynamical critical exponent depending only on the dynamic behavior of the algorithm and on the observable under study. In order to control this problem, we chose to use the \emph{worm algorithm} \cite{ref:1-worm}\cite{ref:2-worm}, a non local algorithm based on the high temperature expansion that drastically reduces $z$. Here we apply this method to the $\phi^4$ theory with $O(N)$ symmetry, whose lagrangian has the general expression
%***
\begin{equation}
\label{eq:Lagr}
	\mathcal{L} =\frac{1}{2}\left(\partial_{\mu}\phi\right)^2 + \frac{\mu_0^2}{2}\left(\phi\cdot\phi\right) + \frac{g}{4}\left(\phi\cdot\phi\right)^2
\end{equation}
%***
where  $\phi$ is a $N$ component field, $\mu_0$ is the bare mass and $g$ is the bare coupling. In particular, in the first part we describe the application of this method to the investigation of the critical features of $\phi^4$ theory in two dimension with $N=1$, showing the upgrades adopted with respect to our previous work \cite{bdpg}. %The convenience of this setting is that the theory is super-renormalisable meaning that the coupling constant has a positive mass dimension and it has only a finite number of 1 Particle Irreducible divergent diagrams. %Furthermore the theories of kind $\phi^4$ are known belonging to the same universality class of the Ising Model and thus we have an enhanced tools for probing this theory. 
Secondly, we present a new version of the worm algorithm, that allows to perform simulation of $\phi^4$ theory with $O(N)$ symmetry in arbitrary dimension. In particular we will show results in the cases with $N=2$, $D=3$. This is a preliminary work that lays the foundation for the investigation of the phase transition of an ideal three-dimensional Bose-Einstein gas with fixed density. The computation of first correction to the shift of the critical temperature \cite{ref:3-Tc}, with arbitrarily weak interactions, is still surrounded by theoretical uncertainties and a way to improve this estimation is to study it by means of an effective field theory. In particular the $O(2)\phi^4_3$ theory on the lattice suits well for probing the non-perturbative features of this physical system.

%////////////////////////////////////////////////////////////////////////////////////////
\section{$\phi^4$ theory in two dimensions}
Now we specialize to the case of $\phi^4$ theory in two dimensions with real fields $\phi$ with $N=1$. From dimensional analysis arguments we know that $[\mu_0^2] = [g]$ and thus the only relevant dimensionless parameter is the ratio $f \equiv g/\mu^2$ defining a critical line when both $g$ and $\mu^2$ goes to zero, where $\mu^2$ now is the renormalised squared mass in some given renormalization scheme. Despite the simplicity of the model there is still debate around this quantity and one of the last Monte Carlo estimation of it is presented in our previous work \cite{bdpg}, obtained by means of the \emph{worm algorithm} technique \cite{ref:2-worm}\footnote{For a summary of the last determinations of $f$ see the Table IV in \cite{bdpg}, with \cite{ref:1} as the most recent estimation.}.  

The formulation of the $\phi^4$ theory on the lattice lead us to the euclidean action
%***
\begin{equation}
\label{eq:S1}
	\mathcal{S} =  \sum_x \left\lbrace -\sum_{\nu}\phi_x\phi_{x+\nu} + \dfrac{1}{2}\left( \mu_0^2+4\right)\phi_x^2 + \dfrac{g}{4}\phi_x^4 \right\rbrace, 
\end{equation}
%***
where $\phi_{x\pm\hat{\nu}}$ are fields at neighbor sites in the $\pm\nu$ directions and $\mu_0$ and $g$ are expressed in lattice units. For the computation of  $f \equiv g/\mu^2$ the  general idea is to fix a value of $g$ and search for a value of $\mu_0^2$ such that we get, in the infinite volume limit, a second order phase transition point in the plane $(g,\mu_0^2)$.
In order to securely get the continuum limit it is necessary to deal with an additive renormalization of the mass parameter, since $\mu^2_0$ diverges like $\log (a)$ in this limit.
%\footnote{Further details can be found in \cite{bdpg}.}. 
In the end we extrapolate the quantity $g/\mu^2$ to the limit $g \to 0$ and finally obtain the critical value in the continuum limit.\\
For our scope is useful to introduce another parametrization of the action:
%***
\begin{equation}
\label{eq:Sbetalambda}
	\begin{split}
	\mathcal{S}_E &= -\beta\sum_x\sum_\nu\varphi_x\varphi_{x+\hat{\nu}} + \sum_x\left[\varphi^2_x + \lambda(\varphi^2_x-1)^2\right] \\
	&= \mathcal{S}_{I} + \mathcal{S}_{Site},
	\end{split}
\end{equation}
%***
where $\mathcal{S}_{I}$ is the interaction term between neighbor sites with a coupling constant of strength $\beta$ and  $\mathcal{S}_{Site}$ is the term related to a single site.
The relations between $(\mu_0^2,\;g)$ and $(\beta,\;\lambda)$ are:
%***
\begin{equation}
\label{param}
	\phi_x = \sqrt{\beta}\varphi, \qquad \mu_0^2 = 2\dfrac{1-2\lambda}{\beta}-4, \qquad g=\dfrac{4\lambda}{\beta^2}.
\end{equation} 
%***
In the next section we briefly describe our computational strategy for the first estimation and for the improved one.
%---
\subsection{Strategy for the computation of $f$}
For the computation of $f$ we use the \emph{worm algorithm}\cite{ref:2-worm} and consider the lattice action  \eqref{eq:Sbetalambda}. 
In order to compute the critical point in the new representation, we fix a value of $\lambda$ and search for the corresponding critical value $\beta$. %o meglio scrivere direttamente $\beta(\lambda)$
 The physical condition we impose for such scope is
 %***
\begin{equation}
\label{eq:mL}
	mL=\dfrac{L}{\xi} = \text{const}= z
\end{equation}
%***
which implies that $\xi$ grows linearly with $L$ and when $L/a\to\infty$ we arrive at the critical point. $m$ is determined by the implicit formula
%***
\begin{equation}
	\dfrac{G(p^*)}{G(0)} = \dfrac{m}{p^{*2} +m^2}
\end{equation}
%***
where $G(p)$ is the two-point function in momentum space, and $p^*$ is the smallest momentum on a lattice of linear size $L$. Basically we simulate several lattices with different values of $N \equiv L/a$; for each couple $(\lambda,\,N)$ we obtain a value of $\beta(\lambda,\,N)$ such that $mL = z$. After this step we extrapolate our results to $a/L\to 0$ in order to compute $\beta(\lambda)$. Now, using relations in \eqref{param} we derive $g(\lambda,\beta)$ and $\mu_{0}^2(\lambda,\beta)$. Using renormalisation condition specified in \cite{bdpg} we finally pin down $\mu^2(g)$ and hence the ratio $g/\mu^2$. We repeat all this procedure for several values of $\lambda$, and hence of $g$; in the end we extrapolate our results to $g\to 0$, in order to get $f$. Our final value is $g/\mu^2 = 11.15 \pm 0.06_{stat} \pm 0.03_{syst}$ obtained imposing $z=4$ in \eqref{eq:mL}. 
%We do not want to go deeper in the details of simulations, for which we refer to \cite{bdpg}, but we only want to focus a little on Table \ref{Tab:tab1} and make some observations. As you can see our result is in good agreement with the last four determinations, excluding the most recent one, and specifically shows a discrepancy at a 3$\sigma$ level with \emph{Monte Carlo cluster} result \cite{L-S}.

\begin{figure}[ht!]
	\begin{center}
		\includegraphics[width=1\textwidth]{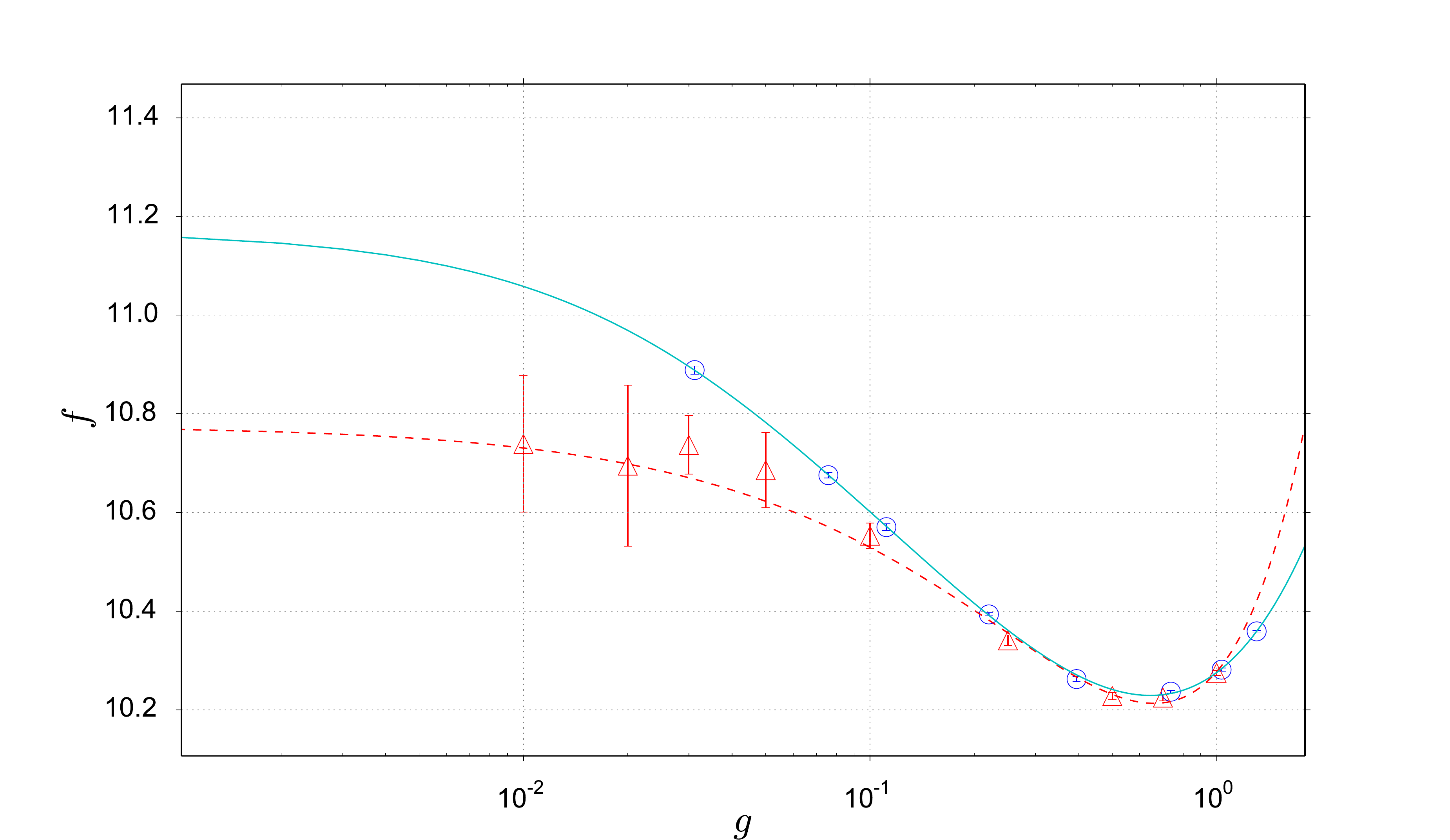}
		\caption{Final results for f(g) in logarithmic scale. The extrapolations are carried out with different methods and with different fit functions.}
		\label{Fig:fig1}
	\end{center}
\end{figure}

In Fig. \ref{Fig:fig1} a direct comparison with the Monte Carlo result obtained with \emph{cluster algorithm} \cite{L-S} is shown: blue round circles are our determinations while red triangular points are the results obtained in \cite{L-S}. In the region of the minimum of the curves the two determinations are in almost perfect agreement, despite the fact that the infinite volume limit extrapolations are obtained with completely different strategy. At lowest values of $g$ our points seem to be a little bit higher but, even if we have more accurate results, we were not able to assert something definitely.
%++++++++++++++++++++++++++++++++++++++++++++++++++++++++++++++++++++++++++++++++++++++++
%++++++++++++++++++++++++++++++++++++++++++++++++++++++++++++++++++++++++++++++++++++++++

In order to improve our estimation we have started a new set of simulations with a slightly different strategy. For several values of $N\equiv L/a$ we perform two independent simulations, following the procedure described above, in which we consider two different values of $z=2,4$. The final extrapolation is obtained by means of a combined fit for the two different choices of $z$.  In this way we do not completely solve the technical problem to probe low $g$-region, but we hope to get more accurate and precise results in order to identify the right behavior of the critical coupling in the continuum limit. %Questo  metodo non risolve il problema tecnico di sondare bassi valori di g ma permett di ottenere valori più accurati in quanto la statistica raddoppia e più precisi in quanto l'estrapolazione è fatta a due condizione di fisica costante diverse. 
%Parametrizzazione adottata per scrivere la nuova lagrangiana (o azione). per maggiori dettagli cit. How we get the continuum limit: we simulate different lattice size $L$ with distinct values $N\equiv L/a$ at fixed coupling $\lambda$; for each couple $(\lambda,N)$ we search a value of $\beta$ such that
%After that we extrapolate our result in order to obtain $\beta(\lambda)$. Then we use the relations (to be written) for deriving $g(\lambda, \beta)$ and $\mu_0^2(\lambda,\beta)$. We finally compute the ratio $f$ using the renormalization condition for $\mu_0^2$ 
%
%Ora il concetto dovrebbe essere che, dato che non riusciamo ad andare a lambda sempre più piccoli e a reticoli sempre più alti, cerhicamo di ottenere informazioni in più utilizzando dati nella regione che riusciamo ad esplorare. In particolare abbiamo eseguito simulazioni con diversi valori di L e lambda per due valori di z=2,4. Dato che l'estrapolazione finale deve portare allo stesso risultato, abbiamo eseguito un fit combinato delle simulazioni. Per ora non abbiamo ancora nessun risultato ma (?) siamo fiduciosi nel futuro. Goal finale?
%Come materiale potrebbe starci qualche tabella di dati.

%////////////////////////////////////////////////////////////////////////////////////////
\section{Extended worm algorithm}
Now we show the extension of the worm algorithm for a $\phi^4$ theory with $N$ component fields $\phi$ and $O(N)$ symmetry. The starting point is the work of U.Wolff \cite{OnWolff} in which the application of the \emph{all-order strong coupling expansion} to the $O(N)$ sigma model is described. 

Consider the partition function with two field insertions, using the action \eqref{eq:Sbetalambda}:
\begin{equation}
\label{PF-ONphi4}
	Z(u,v) = \int \left[\prod_z d\phi(z)\,e^{-S_0}\right]\left(\prod_{l=\langle xy\rangle}e^{\beta\phi(x)\cdot\phi(y)}\right)\phi(u)\cdot\phi(v)
\end{equation}
We want that the term in the squared brackets becomes the new functional measure
\begin{equation}
\label{eq:phi_int_meas}
	\int\prod_x d\phi(x)e^{-|\phi|^2 -\lambda(|\phi|^2 -1)^2} = \int \prod_x d\mu[\phi(x)]
\end{equation}
where the integrations employ the normalized $O(N)$ invariant measure on the sphere, which acts on a test function $f(\phi)$ as follows
\begin{equation}
\int d\mu(\phi)f(\phi) = C_N\int d\,r\,d\theta\,\dfrac{d\,\Omega}{2} r^{N-1}(\sin\theta)^{N-2}f(r,\theta, \Omega).
\end{equation}
$C_N$ is the normalization coefficient, $r$ is the radial integration variable and $\theta, \Omega$  constitute the total solid angle for a $N$-sphere.
Rewriting \eqref{PF-ONphi4} as integral over this new functional measure, we have to work out integration over spherical coordinates in $N$ dimension. In order to obtain the loop representation is useful to introduce the single generating function for a general source $j$
\begin{equation}
\label{eq:G_N}
G_N(j) \equiv \int \prod_x d\mu[\phi(x)] e^{j\cdot\phi} = \sum_{k=0}^{\infty} c[k;N](j\cdot j)^k.
\end{equation}
The coefficients of the series expansion are given by the resolution of the integral by means of the modified Bessel function $I_{N/2-1}$, using the measure defined in \eqref{eq:phi_int_meas}:
\begin{equation}
\label{eq:phi-final}
 c[k;N] = \dfrac{\varrho(N+k-1)\Gamma(N/2)}{\varrho(N-1)2^{2k}k!\Gamma(N/2+k)} .
\end{equation}
With  $\varrho$ we indicate the solution of the radial part of \eqref{eq:G_N} that can be solved only numerically, while the other terms in \eqref{eq:phi-final} come from the resolution of the solid angular part. The \eqref{eq:phi-final} is the key for computing the observables in a $O(N)\phi^4$ model.

\subsection{Algorithm}
The technique \emph{high temperature expansion} allows to translate the original system in a new representation where the new fields are located at links connecting each pair of neighboring sites and have discrete values. Configurations in this formulation have a graphical representation as collections of paths that are called \emph{loops} and the worm algorithm samples them by local moves. In particular there are several update steps that may constitute one local move. Here we only mention those moves that have a different acceptance probabilities with respect to \cite{OnWolff}, implying that the other ones remain the same. For this purpose we need to introduce some definitions: the active loop is the loop participant to the update process, $u$ indicates the head and $v$ the tail of the active loop, and an active loop is called trivial if it contains no 2-vertex and $u = v$. 

Now we write down the ratio $q$ that controls the acceptance probability $P = min(1,q)$ in the cases mentioned before.
\begin{itemize}
	\item Extension: we try to move the head $u$ to one of the nearest neighbor $\tilde{u}$.
	\begin{equation}
	\label{eq:q1}
	q_1 = \dfrac{\varrho(N + d(\tilde{u})/2)}{\varrho(N + d(\tilde{u})/2 -1)}\dfrac{\beta}{N+d(\tilde{u})}
	\end{equation}
	\item Retraction: we try to retract the head $u$ by on link along the active loop and  $\tilde{u}$ is the new head.
	\begin{equation}
	\label{eq:q2}
	q_2 = \dfrac{\varrho(N + d(u)/2-2)}{\varrho(N + d(\tilde{u})/2 -1)}\dfrac{N+d(\tilde{u})-1}{\beta}
	\end{equation}
	\item Kick: if the loop is trivial, we randomly pick a site $x$ and try to move the trivial loop in that site.
	\begin{equation}
	\label{eq:q3}
	q_3 = \dfrac{\varrho(N + d(x)/2)\varrho(N + d(u)/2-2)}{\varrho(N + d(x)/2 -1)\varrho(N + d(u)/2 -1)}\dfrac{N+d(u)-2}{N+d(x)}
	\end{equation}
\end{itemize}
%////////////////////////////////////////////////////////////////////////////////////////
\section{Application}
We perform several simulations to verify our algorithm. In particular we well reproduce the results in the case of the non-linear sigma model for the A-series in \cite{OnWolff}, .  

\begin{figure}[ht!]
	\begin{center}
		\includegraphics[width=.8\textwidth]{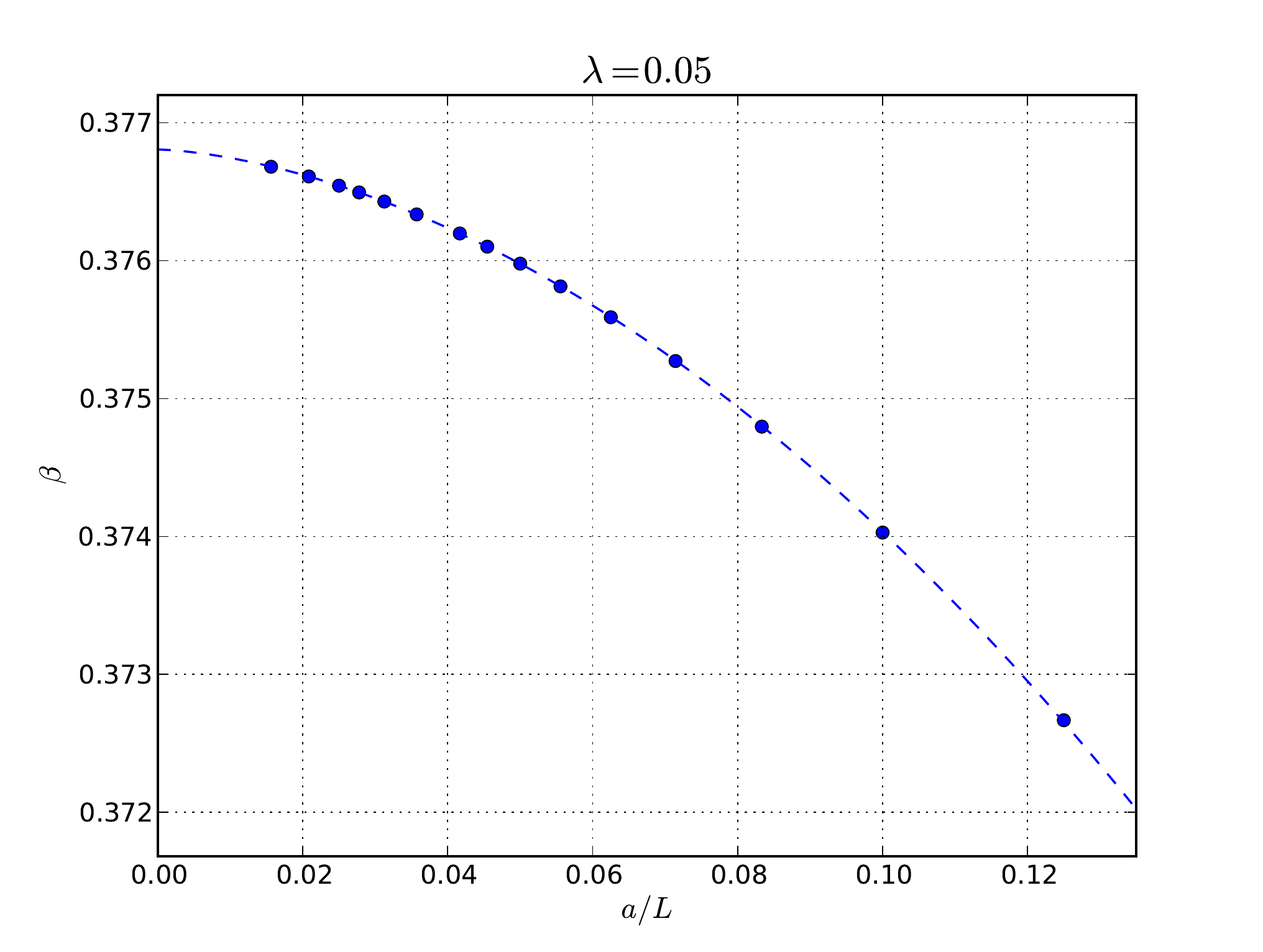}
		\caption{Example of extrapolation of $\beta_c$ in the infinite volume lmit for $\lambda=0.05$.}
		\label{Fig:fig2}
	\end{center}
\end{figure}

We apply our method to the study of the phase transition of the three-dimensional $O(2)\phi^4$, basing on \cite{ref:9}. In Fig \ref{Fig:fig2} we show the extrapolation of $\beta_c$ to the infinite volume limit for $\lambda = 0.05$. 
%++++++++++++++++++++++++++++++++++++++++++++++++++++++++++++++++++++++++++++++++++++++++
%++++++++++++++++++++++++++++++++++++++++++++++++++++++++++++++++++++++++++++++++++++++++
The behavior of the function used for the fit is suggested by finite size scaling arguments and has the following expression
\begin{equation}
\label{eq:beta-ext}
	\beta_{\lambda}(L) = \beta_c + b_1 L^{-1/\nu} +b_2\, L^{-1/\nu +\omega}.
\end{equation}

where the critical exponent $\nu,\,\omega$ are taken from \cite{ref:9}. This is only a preliminary work and here we only want to show the quality of the fit. The next step is the computation of the difference between the value of $\langle \phi^2\rangle$ at the critical point for the case of \textit{i}) small $g$ and \textit{ii}) $g=0$, defined as
\begin{equation}
\label{eq:deltaphi}
\Delta\langle \phi^2\rangle_c \equiv [\langle \phi^2\rangle_c]_g - [\langle \phi^2\rangle_c]_0.
\end{equation}
This is obtained with the same technique adopted in \cite{bdpg}: at fixed $\lambda$ we compute the quantity $\langle \phi^2\rangle$ for several lattice and then we extrapolate in the infinite volume limit. Finally we perform the continuum limit and estimate \eqref{eq:deltaphi}. We do not go further into details since the simulations are ongoing and the results will be presented in a dedicated article.

%////////////////////////////////////////////////////////////////////////////////////////
\section{Conclusion}
In the first part of this paper we summarize the plan we have for improving the estimation of the critical coupling  $g/\mu^2$ in $\phi^4$. Since the simulations are still running, we have not yet a final result. Secondly, we briefly introduce the extension of the worm algorithm for the case of $O(N)\phi^4$ in arbitrary dimensions. In particular we present the application for the case of $D=3$ and $N=2$, summarizing the strategy we adopt for the estimation of \eqref{eq:deltaphi}. 
This quantity is related to the first correction to the shift of the critical temperature of an ideal Bose-Einstein gas and we hope to give soon a new estimation of it. 
%-------------BIBLIOGRAPHY--------------------------------------

\end{document}